%% file: siena.tex
\documentclass[fleqn,twoside]{article}
\usepackage{espcrc2}

\usepackage{graphicx}
\usepackage[figuresright]{rotating}

\newcommand{\AmS}{{\protect\the\textfont2
  A\kern-.1667em\lower.5ex\hbox{M}\kern-.125emS}}

\input{definitions}

\title{The physics goals of the TESLA project}

\author{K. Moenig\address{DESY-Zeuthen}%
}
       
\begin{document}

\begin{titlepage}

\pagenumbering{arabic}
\begin{flushright}
hep-ex/0112001\\
3rd December 2001
\end{flushright}
\vspace*{2.cm}
\begin{center}
\Large 
\boldmath
{\bf
The physics goals of the TESLA project
} \\
\unboldmath
\vspace*{2.cm}
\normalsize { 
   {\bf K. M\"onig}\\
   {\footnotesize DESY-Zeuthen}\\
   
}
\end{center}
\vspace{\fill}
\noindent
As next generation $\ee$ linear collider the superconducting accelerator
project TESLA has been proposed. In this note the physics potential goals of 
this project, which is highly complementary to LHC, are described.

\vfill
\noindent
Invited talk presented at the 
``Seventh Topical Seminar on The legacy of LEP and SLC'', Siena, October 2001
\vspace{\fill}
\end{titlepage}
%
%
%
\setcounter{page}{1}    

\begin{abstract}
As next generation $\ee$ linear collider the superconducting accelerator
project TESLA has been proposed. In this note the physics potential goals of 
this project, which is highly complementary to LHC, are described.
\end{abstract}

\maketitle

\section{INTRODUCTION}
One of the most important questions at the next generation high energy 
colliders is how the electroweak symmetry is broken. With the present knowledge
several ways are considered to be possible.
The electroweak precision data indicate that probably masses are generated by
a light Higgs boson. Although by theoretical arguments it is very improbable,
in this case the Standard Model could be the final theory or at least look
like that for the next generation of colliders.
More probable, however, new physics is close by, where the currently
discussed scenarios are mainly Supersymmetry or extra space dimensions.
As a second possibility no elementary scalar exists but the Higgs mechanism
is mimicked by a new strong interaction.

In all cases it is likely that the LHC, if not already the Tevatron, will see
first signals of the mechanism at work. However also in all cases complementary
information from a lepton collider is needed to understand the underlying 
theory.

At present three designs for a next generation $\ee$ collider are under study:
TESLA in Europe \cite{tdrmachine}, NLC is the US \cite{nlc} and 
JLC in Japan \cite{jlc}. All three designs cover the 
energy range up to around 1\,TeV and could be ready around 2010. This report
concentrates on TESLA, but the differences in the physics potential of
the different machines are minor.

In its first stage TESLA can run at centre of mass energies between the Z-pole
and 500\,GeV. An upgrade to close to 1\,TeV is planned. The predicted 
luminosities vary between $50 \fbi$/year at the Z-pole and $500 \fbi$/year
at 800\,GeV. This corresponds to around one billion Zs per year on the Z-pole,
sixty thousand Higgses/year at $\sqrt{s}=350 \GeV$ ($\MH = 120 \GeV)$,
a hundred thousand top pairs per year at the peak of the cross section
or a million W-pairs/year at higher energies.

In March 2001 the technical design report for TESLA was presented including
a detailed discussion of the physics case for a linear 
collider \cite{tdrphys}. If not explicitly mentioned otherwise all
studies presented in this note are taken from there.
Other topics, like extra dimensions, extended gauge theories or QCD studies, 
not mentioned here, can also be found in \cite{tdrphys}. 

\section{TOP QUARK PHYSICS}
The linear collider will be the first opportunity for a detailed exploration 
of the top threshold. Top physics is interesting for several reasons.
It may not be a pure accident that the top quark mass is close to the vacuum 
expectation value of the Higgs and a future theory of flavour might need
accurate measurements in the quark sector.
However, the top quark enters also the radiative corrections to other processes
and especially the knowledge of $\MT$ is needed not to obscure
the value of the electroweak precision measurements. 
The contribution of the top quark mass to
the effective weak mixing angle is $\Delta \stl / \Delta \MT = 0.00003/\GeV$
requiring a top mass precision of a few hundred MeV for the ultimate 
precision of $\Delta \stl \sim 0.00002$ from TESLA. In Supersymmetry 
the influence of $\MT$ on the Higgs mass is 
$\Delta \MH / \Delta \MT \sim 1$ requiring a precision of the top quark mass
below 100\,MeV.

Experimentally $\MT$ can be measured in a threshold scan to a precision below
50\,MeV. With the new mass definitions and next to next to leading log
calculations also the theoretical understanding of the top mass improved
enormously \cite{mttheo}, so that a total top mass precision in the
$\overline{\rm MS}$ scheme of 100-200\,MeV is within reach.
In addition to the mass measurement several other top quark properties can 
be measured. As one example the top quark width can be obtained from
the shape of the threshold curve and the forward backward asymmetry slightly
above threshold with a few percent precision.

\section{HIGGS PHYSICS}
Theoretical arguments as well as the electroweak precision data point
towards a light Higgs, well within reach of the first stage of TESLA.  
If it exists, its exploration will be 
one of the most important subjects of linear collider physics.

If a light Higgs exists almost certainly it will be discovered by the Tevatron
or the LHC before the start of a linear collider. 
The task of TESLA is then to verify that
the particle compatible with a Higgs is really the particle responsible for
mass generation and to do precision measurements of the Higgs to
verify that the Higgs sector is as predicted in the Standard Model or,
if deviations are found, to
estimate parameters in an extended Higgs sector.
Higgs-bosons are produced at TESLA in two processes, Higgsstrahlung 
$(\ee \rightarrow {\rm ZH})$ and vector-boson fusion 
($\ee \rightarrow {\rm WW}\nu\nu \rightarrow {\rm H}\nu\nu$,
$\ee \rightarrow {\rm ZZ}\ee \rightarrow {\rm H}\ee$), where ZZ fusion is
much smaller than WW fusion. 
For $\MH =120 \GeV$ and $\sqrt{s}\approx 350 \GeV$ Higgstrahlung dominates with
$\sigma \approx 150 \fb$ over the fusion process with a cross section of
about $30 \fb$.
For $\MH \sim 500 \GeV$ and $\sqrt{s}\approx 800 \GeV$ both processes have 
cross sections of ${\cal O}(10 \fb)$.

\begin{figure}[htb]
\includegraphics[width=\linewidth,bb=0 20 567 559]{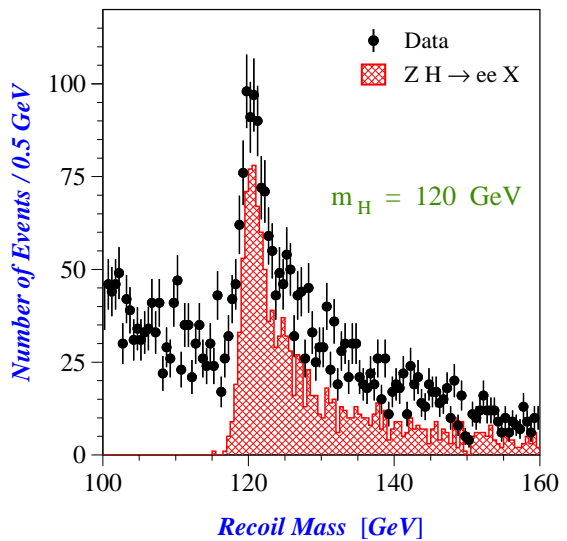}
\vspace{-1cm}
\caption{$\ee$-recoil mass of $\ee \rightarrow \ee X$ events together with
the $\ee \rightarrow {\rm ZH}$ signal.
}
\label{fig:zhrec} 
\end{figure}

If a particle compatible with a Higgs is found, it has to be verified
that its spin and parity are really $0^+$. This can be done unambiguously
showing in a threshold scan that the cross section rises with $\beta$ and
and measuring the angular distribution of the $\rm Z \rightarrow ZH$-decay.
Another extremely important analysis is the measurement
of the total Higgsstrahlung cross section. This cross section measures
directly the ZH coupling testing if the seen Higgs boson is fully responsible
for the Z mass. In addition it gives an absolute normalisation for the 
Higgs branching ratio measurements. 
The cross section measurement can be done in a completely model independent 
way looking at
the Z-recoil mass distribution in $\ee \rightarrow {\rm ZX}$ with 
${\rm Z} \rightarrow \ee,\mumu$. Figure \ref{fig:zhrec} shows this
recoil mass
distribution for $\MH = 120 \GeV$ and $\sqrt{s} = 350 \GeV$. In one to two
years of running a precision of $2.5 \%$ can be achieved.

\begin{figure}[htb]
\begin{center}
\includegraphics[width=\linewidth]{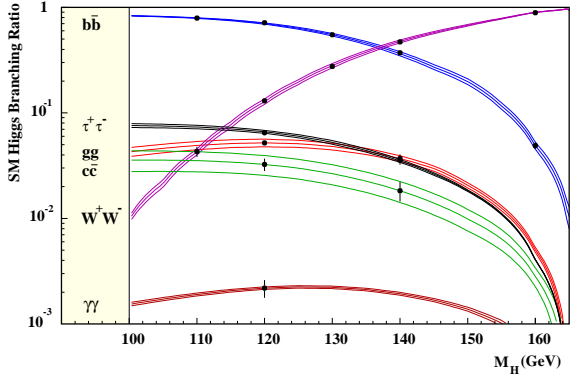}
\end{center}
\vspace{-1cm}
\caption{Predicted Higgs branching ratios and possible error of the
Higgs branching ratio measurements with TESLA as a function of the
Higgs mass.
}
\label{fig:hbr} 
\end{figure}

With the excellent flavour tagging and energy flow resolution of the proposed
detector \cite{paolo} many decay modes of the Higgs can be measured. 
Figure \ref{fig:hbr}
shows the predicted branching ratios and the estimated uncertainties for
these decays as a function of the Higgs mass. For example for $\MH=120\GeV$
one can measure ${\rm BR(H} \rightarrow \bb)$ with $2\%$,
${\rm BR(H} \rightarrow \cc)$ with $8\%$,
${\rm BR(H} \rightarrow gg)$ with $6\%$,
${\rm BR(H} \rightarrow \tau^+ \tau^-)$ with $5\%$,
${\rm BR(H} \rightarrow \WW)$ with $5\%$ and
${\rm BR(H} \rightarrow \gamma\gamma)$ with $20\%$ precision.

\begin{figure}[htb]
\begin{center}
\includegraphics[width=\linewidth]{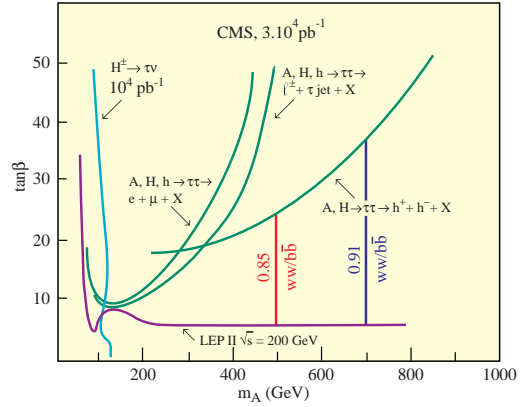}
\end{center}
\vspace{-1cm}
\caption{Predicted sensitivity to $\MA$ as a function of $\tan \beta$ for
the direct searches at LHC and the Higgs branching ratio measurement at TESLA.
}
\label{fig:hbrma} 
\end{figure}

Within the MSSM the ratio of the $\WW$ and the $\bb$ branching fraction
of the Higgs is sensitive to $\MA$, independent of $\tan \beta$. With the
experimental errors given above the sensitivity extends to A-masses up to
500-700 GeV as can bee seen from figure \ref{fig:hbrma} \cite{francois}. 
Also the fusion process 
$\ee \rightarrow {\rm WW}\nu\nu \rightarrow {\rm H}\nu\nu$ which is sensitive
to $\Gamma({\rm H \rightarrow WW})$ can be measured with 3\% accuracy
for $\MH = 120 \GeV$, so
that the total width of the Higgs can be measured in a completely model
independent way with a precision of 6\% when combined with the branching
ratio ${\rm H \rightarrow WW}$.

The Higgs is probably too light that it can decay into a pair of top-quarks.
However at $\sqrt{s} = 800 \GeV$ the top-Higgs Yukawa coupling can be measured
from the rate of events of the type $\ee \rightarrow \ttb {\rm H}$
where a Higgs is radiated off a top quark to a precision of around 5\%
for $\MH = 120 \GeV$.

\begin{figure}[htb]
\begin{center}
\includegraphics[width=\linewidth]{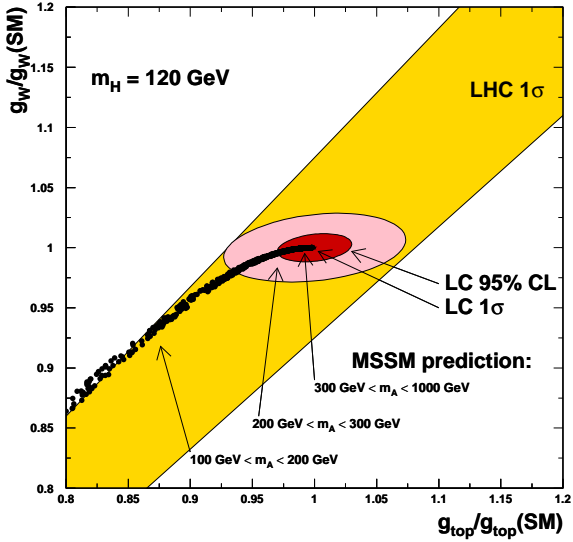}
\end{center}
\vspace{-1cm}
\caption{Predicted accuracy for the Higgs couplings to W-bosons and
  top quarks at TESLA and LHC.
}
\label{fig:hcoup} 
\end{figure}

With the Higgs cross sections and branching ratios measured at TESLA
the Higgs couplings can be determined. As an example figure \ref{fig:hcoup}
shows the predicted accuracy for the Higgs couplings to the W-boson
and the top-quark for TESLA and the LHC.

Another very important measurement is the reconstruction of the Higgs
potential. In the SM the potential is predicted to be 
$V(\Phi)= \lambda (\Phi^* \Phi -v^2/2)^2$, where $v$ is fixed by the
muon lifetime and $\lambda$ by the Higgs mass. The trilinear and
quartic Higgs self-couplings are thus given by the shape of the Higgs potential.

The quartic Higgs-coupling is much too small to be measured at a next
generation machine. However the triple Higgs coupling contribution to
the $\ee \rightarrow {\rm ZHH}$ cross section is large enough that it can
be measured. However, similar to $\ttb {\rm H}$ the measurement of this
process requires a very good detector performance. The cross section
of a few fb has to be filtered out of a potentially huge background of
multi-jet final states. To do this one needs very good flavour
tagging capabilities to identify the four b-quarks in the final state
and very good energy flow resolution in hadronic jets to reconstruct
the intermediate resonances. However, if the proposed performances can
be reached a measurement of the triple-Higgs coupling with about 20\%
accuracy seems possible.

\section{SUPERSYMMETRY}

The most popular extension of the Standard Model is certainly Supersymmetry
(SUSY). SUSY does not only solve the hierarchy problem but
it enables grand unification at a high scale
and, if r-parity is conserved, it offers
also an excellent candidate for dark matter. 
There are many arguments that some supersymmetric particles, if they exist, 
should be in the energy range of TESLA.

\begin{figure}[htb]
\begin{center}
\includegraphics[width=0.48\linewidth]{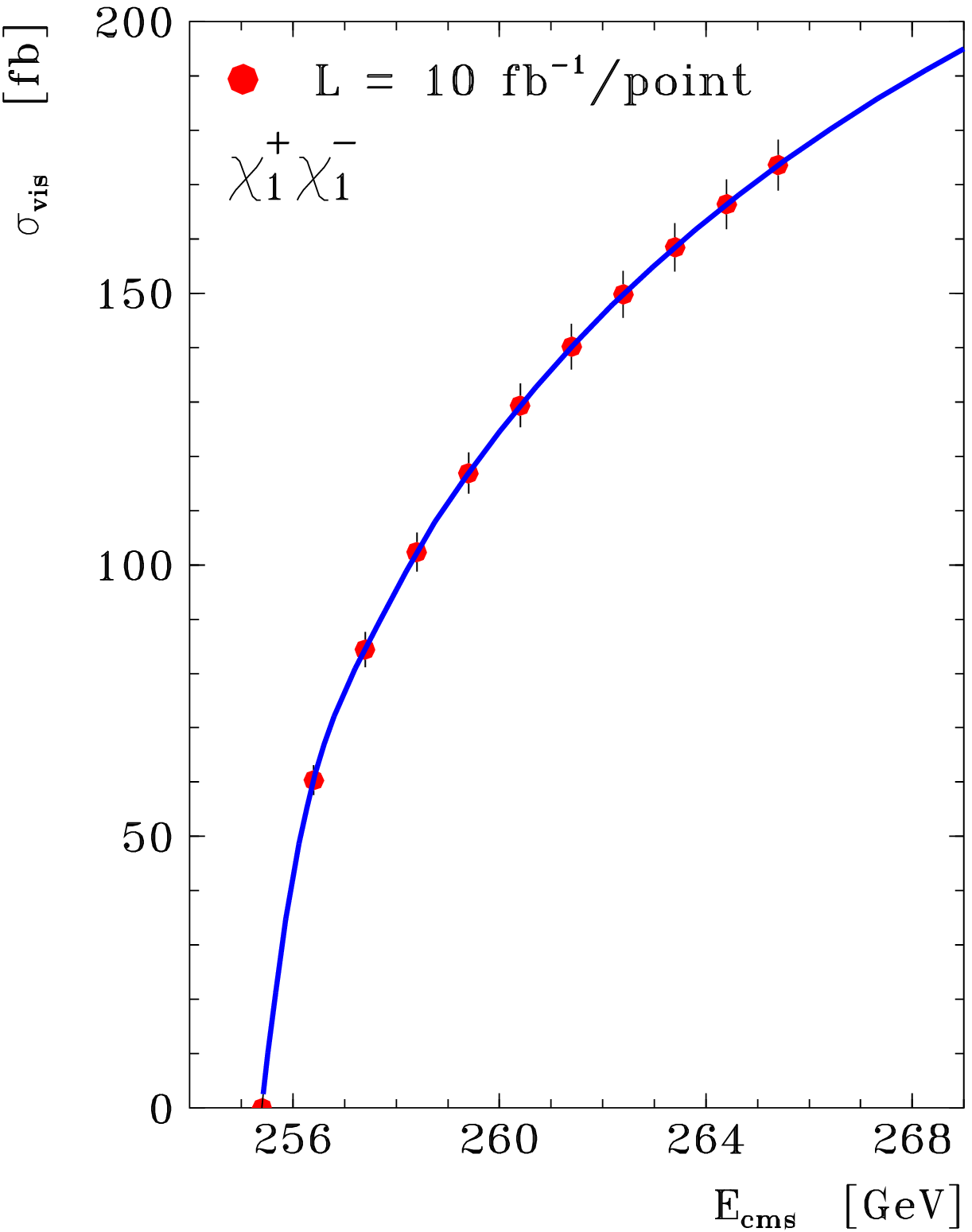}
\includegraphics[width=0.48\linewidth]{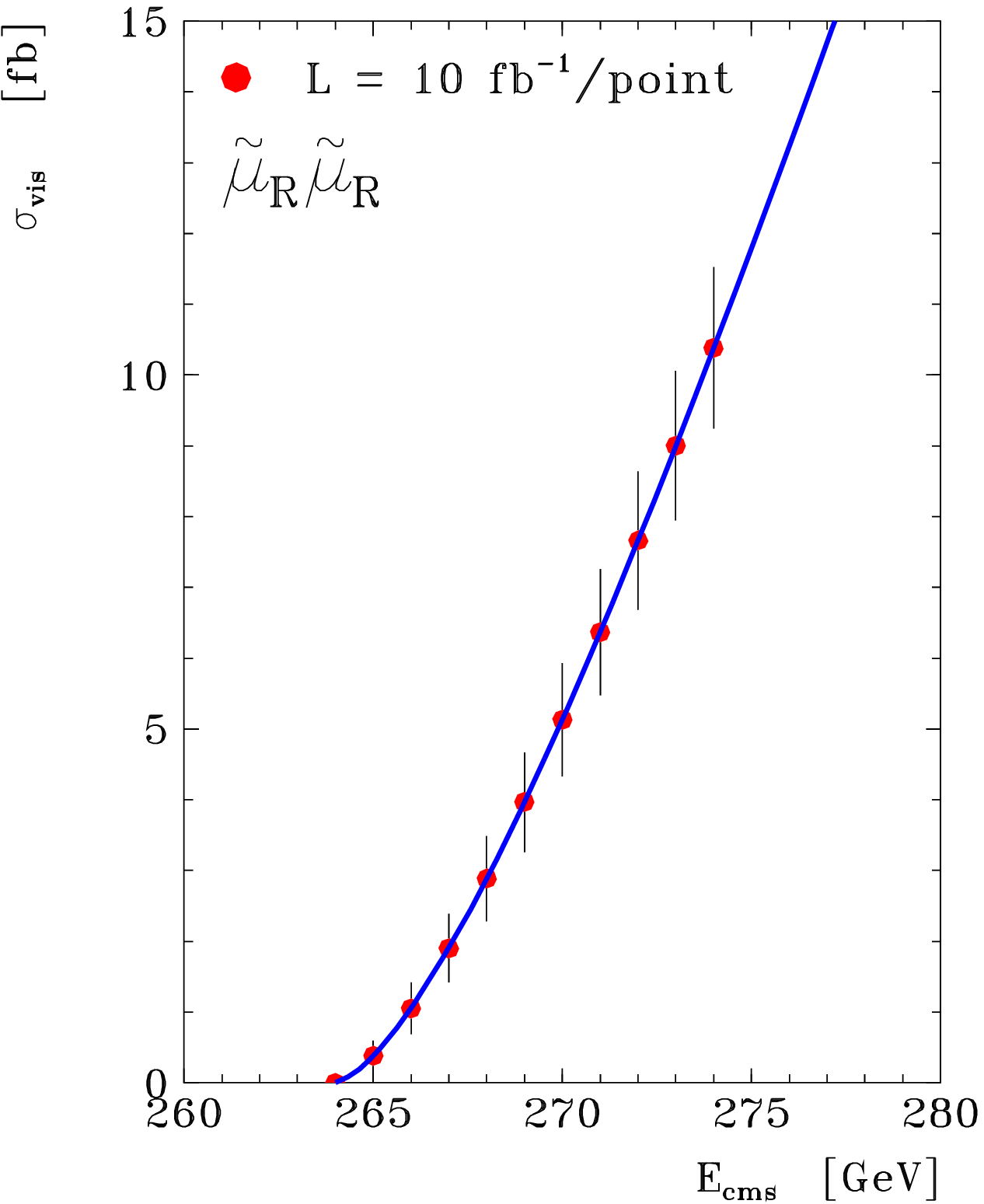}
\end{center}
\vspace{-1cm}
\caption{Simulated threshold scan for $\ee \rightarrow \chi_1^+ \chi_1^-$ 
  and $\ee \rightarrow \tilde{\mu}_R^+ \tilde{\mu}_R^0$.
}
\label{fig:susythresh} 
\end{figure}

If SUSY exists it has to be broken, leading to more than 100 new free
parameters. The most important task, once SUSY is discovered, is thus
to measure as many of the new parameters as possible to understand the
mechanism of SUSY breaking. The best way to access these parameters is the
measurement of the masses of the SUSY particles. This can be done mainly
in two ways, either performing a threshold scan or measuring the masses
reconstructing the final state particles.
For the production of pairs of identical charginos or neutralinos the 
threshold suppression is proportional to
$\beta$. With this steep rise of the cross section e.g. the
$\chi_1^\pm$ mass can be measured to $0.05\%$ from a threshold scan. 
For sfermions the threshold
suppression is $\propto \beta^3$, making the precision a factor two
worse. Figure \ref{fig:susythresh} shows the threshold cross section for
chargino- and right handed smuon pairs.

\begin{figure}[htb]
\begin{center}
\includegraphics[width=\linewidth]{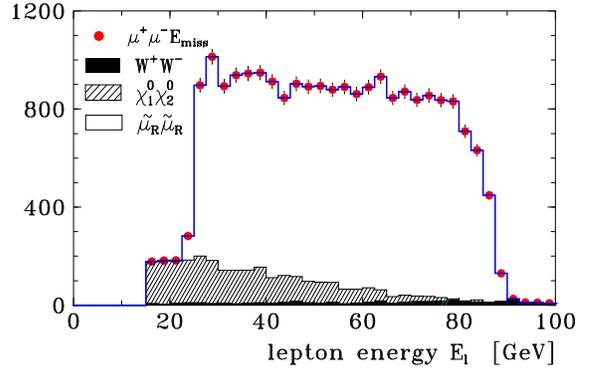}
\end{center}
\vspace{-1cm}
\caption{Simulated muon energy spectrum for the process
$\ee \rightarrow \tilde{\mu}_R^+ \tilde{\mu}_R^- \rightarrow
\mu^+ \chi^0_1 \mu^- \chi^0_1$ ($\sqrt{s}=320 \GeV,\,{\cal L}=160 \fbi,\,
m_{\tilde{\mu}_R}=132 \GeV,\,m_{\chi^0_1}=72\GeV $).
}
\label{fig:susymspec} 
\end{figure}

Mass reconstruction is most precise for the sfermion decay 
$\tilde{f} \rightarrow f \chi$. Since the sfermions are scalar, the fermion
energy distribution has to be flat between 
${E_f}/{E_{\rm{beam}}} \, = \, {1}/{2} \left( 1 \pm \beta \right)
\left( 1 - ({m_\chi}/{m_{\tilde{f}}})^2 \right)$, 
from which 
$m_{\tilde{f}}$ and $m_\chi$ can be reconstructed in a model independent way. 
As an example of this method, figure \ref{fig:susymspec} shows the muon energy
spectrum for the process 
$\ee \rightarrow \tilde{\mu}_R^+ \tilde{\mu}_R^- \rightarrow
\mu^+ \chi^0_1 \mu^- \chi^0_1$. The events can be selected cleanly and
$m_{\chi^0_1}$ and $m_{\tilde{\mu}_R}$ can be measured with an accuracy of
$0.3 \%$ with $\sqrt{s}=320 \GeV,\,{\cal L}=160 \fbi$.

\begin{figure}[htb]
\begin{center}
\includegraphics[width=0.8\linewidth,bb=17 3 468 460]{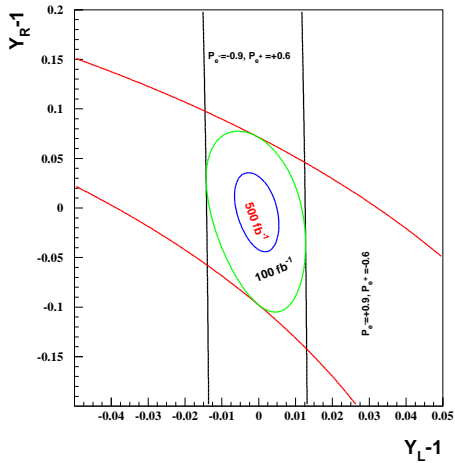}
\end{center}
\vspace{-1cm}
\caption{Possible precision on
the left- and right-handed selectron coupling to neutralinos relative to
the prediction using neutralino pair production
with polarised beams.
}
\label{fig:susycoup}
\end{figure}

If not all particles are accessible or if, because of CP-violating phases,
not all parameters can be measured from the masses alone, additional 
sensitivity to the SUSY breaking parameters can be gained from
the measurement of
the polarised cross section. Even if all parameters are known from the
masses, these cross section measurements are still very important to measure
the couplings of the SUSY particles and to show that they are really
identical to the ones of their Standard Model partners.
Figure \ref{fig:susycoup} shows the precision that can obtained on
the left- and right-handed selectron coupling to neutralinos relative to
the prediction assuming identical couplings for Standard Model particles
and their superpartners. For this plot only neutralino pair production
with polarised beams
has been used, where for the gaugino component the relevant production
process is selectron t-channel exchange. With $500 \fbi$ the left
handed coupling can be measured well below 1\% and the right handed coupling
below 5\% \cite{gudi}.

\begin{figure*}[htb]
\begin{center}
\includegraphics[width=0.8\linewidth]{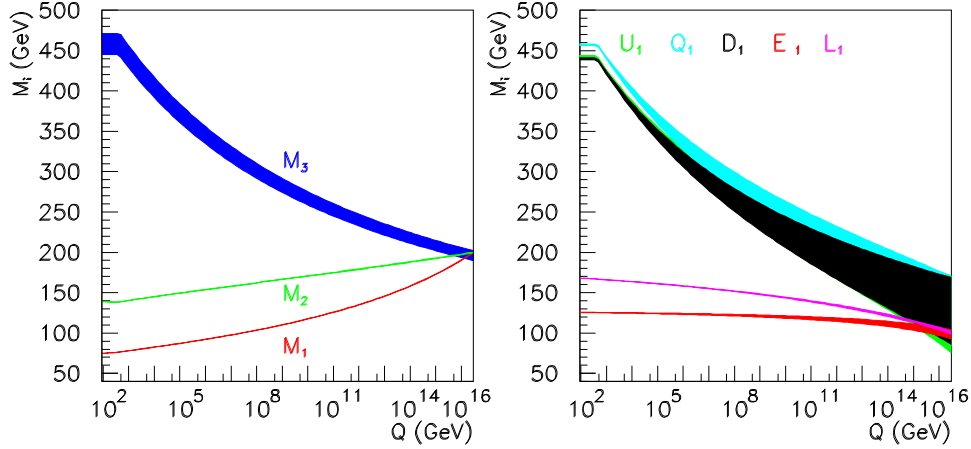}
\end{center}
\vspace{-1cm}
\caption{Extrapolation of a possible SUSY parameter set from the weak scale
to $m_{\rm GUT}$.
}
\label{fig:susygut}
\end{figure*}

If the SUSY parameters are known at the electroweak scale, they can be
extrapolated to higher scales using the renormalisation group
equations in an almost model independent way. This is shown in figure
\ref{fig:susygut} for a parameter set taken from an MSUGRA model. It can
then be checked experimentally if the data support grand unification.

For studies of supersymmetric theories the LHC and TESLA are highly
complementary. At the LHC SUSY events start normally with squarks or gluinos
having masses up to 2\,TeV. Depending on the parameters, sleptons and
gauginos might be seen in cascade decays.
At TESLA especially sleptons and gauginos can be measured very well.
Squarks and gluinos are probably too heavy and in addition the gluino
cross section is extremely small. 
For the particles accessible at the linear collider TESLA has in general
the much better precision. The LHC can measure pretty accurate mass
differences. However, to convert these differences into absolute squark and 
gluino masses a $m_{\chi^0_1}$ measurement from a linear collider
would be needed.

Since the two machines access different particles it cannot be said in
general which one has the higher reach. In most models the LHC reach turns
out to be higher, but also models can be constructed, where TESLA sees
superpartners and the LHC does not.

\section{STRONG ELECTROWEAK SYMMETRY BREAKING}

\begin{figure}[htb]
\begin{center}
\includegraphics[width=0.6\linewidth]{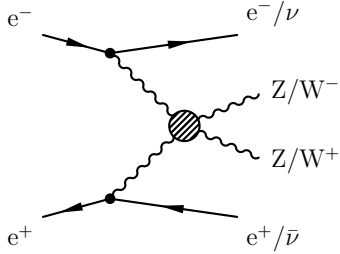}
\end{center}
\vspace{-1cm}
\caption{Generic Feynman diagram of VV scattering in $\ee$.}
\label{fig:feyn_wwscat}
\end{figure}

If no light Higgs exists electroweak interactions become strong at high
energies. The process $\WW \rightarrow \WW$ violates unitarity at 
$\sqrt{s} = 1.2 \TeV$ in this case. At the latest at this energy new physics
has thus to set in to regularise the WW cross section. According to
the low energy equivalence theorem the scattering of longitudinal vector
bosons can be related to pion scattering in QCD and most models for strong
electroweak symmetry braking predict new resonances, corresponding to
the $\rho, \, \omega$ etc.~in QCD, decaying into gauge bosons.
WW scattering can be measured at TESLA with the process depicted in figure
\ref{fig:feyn_wwscat}. However, since the energy of the radiated Ws is 
on average much lower than the one of the radiating electron the sensitivity
of this process is limited. Nevertheless, a new physics scale of
around 3\,TeV can be probed at TESLA with $\sqrt{s}=800\GeV$, which is about
the maximum expected from unitarity arguments.

\begin{figure}[htb]
\includegraphics[width=\linewidth]{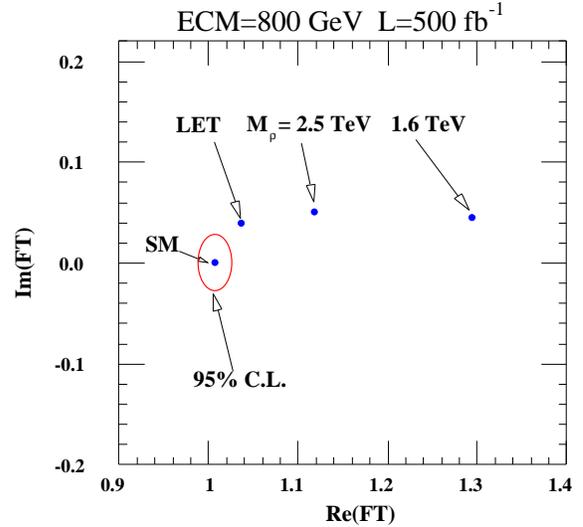}
\vspace{-1cm}
\caption{Sensitivity of $\ee \rightarrow \WW$ to techni-$\rho$ production.}
\label{fig:wwtrho} 
\end{figure}

However in the same way, as the $\rho$-meson is seen 
$\ee \rightarrow \pi^+ \pi^-$ the new physics responsible for electroweak
symmetry breaking should be visible in $\ee \rightarrow \WW$.
Figure \ref{fig:wwtrho} shows a model dependent analysis, where the 
$\ee \rightarrow {\rm W}^+_L {\rm W}^-_L $ amplitude is multiplied by
a form factor depending on the mass of the $\rho$-like resonance. 
It can be seen that TESLA is sensitive to $\rho$-masses of around
2.5\,TeV and can distinguish clearly between the Standard Model and
the prediction of the equivalence theorem.

\begin{figure}[htb]
\includegraphics[width=\linewidth,bb=35 37 495 473]{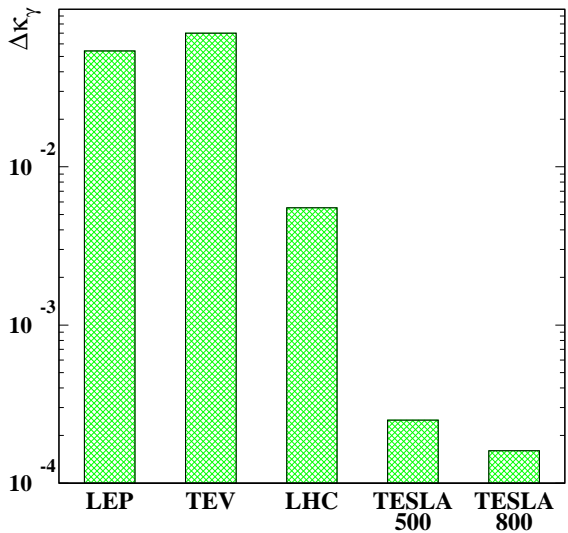}
\vspace{-1cm}
\caption{Sensitivity of various machines to $\kappa_\gamma$.
}
\label{fig:tgccomp} 
\end{figure}

In a more systematic approach the process $\ee \rightarrow \WW$ 
can be parameterised
with an effective Lagrangian, where three terms,
$\alpha_1,\, \alpha_2,\, \alpha_3$, violating custodial SU(2)
at most linearly, affect the triple gauge couplings. These terms can be
written in terms of $g_1^Z,\, \kappa_\gamma,\, \kappa_Z$ which are
already measured at LEP \cite{gideon}. As can be seen from figure 
\ref{fig:tgccomp} TESLA can measure these
parameters to a precision around $10^{-4}$, far superior to any other machine.
The WWZ and WW$\gamma$ couplings can be separated using beam polarisation.

The transformation form $g_1^Z,\, \kappa_\gamma,\, \kappa_Z$ to
$\alpha_1,\, \alpha_2,\, \alpha_3$ is singular, so that only two of
the three $\alpha$-parameters can be measured from W-pair production.
Fortunately $\alpha_3$ is tightly constrained by loop corrections to
the weak mixing angle and the W-mass, so that this degeneracy only slightly
affects the precision on the $\alpha$-parameters, 
if $\stl$ is measured to around 0.00002 and $\MW$ to 6\,MeV by TESLA.
In this case the expected precision for the $\alpha_i$ corresponds to a
new physics scale around 10\,TeV which is much larger than the 3\,TeV
expected from unitarity arguments so that W-pair production should
give first signs of strong electroweak symmetry breaking if no light
Higgs exists.

\section{CONCLUSIONS}

TESLA, with an energy reach of up to around 1\,TeV and an integrated
luminosity of several hundred $\fbi$/year has the potential to enlarge
our knowledge on electroweak symmetry breaking significantly.
If a light Higgs exists, especially in conjunction with Supersymmetry, a
huge amount of information can be obtained from such a machine.
However also in the case without a light elementary scalar 
first hints towards a theory of strongly interacting symmetry breaking can 
be obtained.

In all cases TESLA is complementary to the LHC and we 
finally need both of them to understand electroweak symmetry breaking.
The physics case is strong enough to approve a linear collider project now
and also the LHC would profit from simultaneous running with TESLA.

\section*{ACKNOWLEDGEMENTS}
I would like to thank the organisers for the splendid organisation of the
conference. Also I thank Klaus Desch for reading the manuscript.

\newcommand{\href}[1] {\typeout{#1}}

\begingroup\raggedright\endgroup

\end{document}

%% file: definitions.tex
\def\ifmath#1{\relax\ifmmode #1\else $#1$\fi}%

\def\TeV{\ifmmode {\,\mathrm{ Te\kern -0.1em V}}\else
                   \textrm{Te\kern -0.1em V}\fi}%
\def\GeV{\ifmmode {\,\mathrm{ Ge\kern -0.1em V}}\else
                   \textrm{Ge\kern -0.1em V}\fi}%
\def\MeV{\ifmmode {\,\mathrm{ Me\kern -0.1em V}}\else
                   \textrm{Me\kern -0.1em V}\fi}%
\def\keV{\ifmmode {\,\mathrm{ ke\kern -0.1em V}}\else
                   \textrm{ke\kern -0.1em V}\fi}%
\def\eV{\ifmmode  {\,\mathrm{ e\kern -0.1em V}}\else
                   \textrm{e\kern -0.1em V}\fi}%

\newcommand{\fb}{\,\rm{fb}}

\newcommand{\fbi}{\,\rm{fb}^{-1}}

\newcommand{\WW}    {\mathrm{W}^+\mathrm{W}^-}
\newcommand{\ee}    {\mathrm{e}^+\mathrm{e}^-}
\newcommand{\mumu}  {\mu^+\mu^-}
\newcommand{\ttb}   {{\mathrm t\bar{\mathrm t}}}
\newcommand{\bb}    {{\mathrm b\bar{\mathrm b}}}
\newcommand{\cc}    {{\rm c\bar{\rm c}}}

\newcommand{\MH}      {m_{\mathrm{H}}}

\newcommand{\MA}      {m_{\mathrm{A}}}

\newcommand{\MW}      {m_{\mathrm{W}}}
\newcommand{\MT}      {m_{\mathrm{t}}}

\newcommand {\stl}  {\sin^2 \theta_{e\!f\!f}^l}
